\documentclass[12pt]{article}
\usepackage{amssymb,amsmath,epsfig}

\begin{document}

\title{\bf Thermodynamics in Kaluza-Klein Universe}
\author{M. Sharif \thanks {msharif.math@pu.edu.pk} and Rabia Saleem
\thanks{rabiasaleem1988@yahoo.com}\\
Department of Mathematics, University of the Punjab,\\
Quaid-e-Azam Campus, Lahore-54590, Pakistan.}

\date{}
\maketitle

\begin{abstract}
This paper is devoted to check the validity of laws of
thermodynamics for Kaluza-Klein universe in the state of thermal
equilibrium, composed of dark matter and dark energy. The
generalized holographic dark energy and generalized Ricci dark
energy models are considered here. It is proved that the first and
generalized second law of thermodynamics are valid on the apparent
horizon for both of these models. Further, we take a horizon of
radius $L$ with modified holographic or Ricci dark energy. We
conclude that these models do not obey the first and generalized
second law of thermodynamics on the horizon of fixed radius $L$ for
a specific range of model parameters.
\end{abstract}

\section{Introduction}

The well-established notion is that the universe has entered in the
phase of accelerating expansion. Type Ia supernovae
\cite{1}-\cite{3}, cosmic microwave background radiation (CMBR)
\cite{4}, Wilkinson microwave anisotropy probe (WMAP) \cite{5} and
Sloan digital sky survey (SDSS) \cite{6,7} has indicated that our
universe is flat, homogeneous and isotropic over large scale. This
speedy expansion of our universe is due to an antigravity force
which is drawing galaxies apart from each other, dubbed as dark
energy (DE). Some scientists believe that extra dimensions of space
are also responsible for this expansion. The mechanism behind this
expansion and the nature of DE is not very much clear. Dark energy
having large negative pressure dominates $76\%$ energy density of
the universe \cite{8}.

The cosmological constant is the most suitable candidate of DE which
may be characterized by an equation of state (EoS) parameter,
$\omega=-1$. The current value of this constant is $10^{-55}cm^{-2}$
whereas in particle physics it is $10^{120}$ times greater than this
factor, this problem is known as fine-tuning problem \cite{9}. The
other serious problem is the cosmic coincidence problem which raised
due to the comparison of dark matter (DM) and DE in the present
expanding universe. There have been many DE models proposed such as
scalar field models and interacting models etc. Quitessence
\cite{10,11}, k-essence \cite{12}, phantom \cite{13,14}, tachyon
\cite{15,16}, and quintom \cite{17,18} are the scalar field models
while the interacting DE models are Chaplygin gas \cite{19,20},
braneworld \cite{21,22} and holographic DE (HDE) \cite{23,24}.
Unfortunately, this whole class of DE models do not explain the
nature and its origin in a comprehensive way.

According to recent observations, multidimensional theories may help
to resolve such problems of cosmology and astrophysics. The most
impressing theory in this scenario is offered firstly by Kaluza
\cite{25} and Klein \cite{26} by adding an extra dimension in
general relativity (GR), known as Kaluza-Klein (KK) theory. It is
basically a five dimensional (5D) theory in which gravity is unified
with electromagnetism through this extra dimension. The validity of
laws of thermodynamics has been discussed with modified HDE (MHDE)
\cite{27}-\cite{31}. Some authors \cite{32}-\cite{35} extended this
work to modified gravity theories like $f(R),~f(T)$, Brans-Dicke
(BD) and Horava-Lifshitz theory. Sharif and Khanum \cite{36} checked
the validity of generalized second law of thermodynamics (GSLT) in
KK universe with interacting MHDE and DM. Recently, Sharif and Jawad
\cite{37} explored this work with varying $G$ to investigate the
validity of GSLT in the same scenario.

Holographic DE model based on the holographic principle, is a good
effort in quantum gravity to understand the nature of DE to some
extent. According to this principle, a physical system placed inside
a spatial region is observed with its area but not within its volume
\cite{38}. Cohen et al. \cite{39} argued the cosmological version of
this principle, the quantum zero-point energy $(\rho_\Lambda)$ of
the system having size $L$ (infrared cutoff) cannot exceed the mass
of a black hole (BH) with the same size. Mathematically, we get an
inequality i.e., $L^3\rho_{\Lambda}\leq LM_{p}^{2}$, where $M_p$ is
the reduced Planck mass expressed as $M_p=(8\pi G)^{-\frac{1}{2}}$.
This inequality is most suitable for large $L$ with event horizon.
The HDE density can be expressed as
$\rho_\Lambda=3c^2M_p^{2}L^{-2}$, where $3c^2$ is a dimensionless
constant. The HDE in modified version for KK theory is known as MHDE
\cite{40} and can be calculated from the $(N+1)$-dimensional mass of
the BH \cite{41}.

Ricci DE (RDE) \cite{42} is a type of DE obtained by taking square
root of the inverse Ricci scalar as its infrared cutoff. Gao et al.
\cite{43} explored that the DE is proportional to the Ricci scalar.
Some recent work \cite{44}-\cite{46} shows that the RDE model fits
well with observational data. Xu et al. \cite{47} gave the
generalization of two dynamical DE models, i.e., generalized HDE
(GHDE) and generalized RDE (GRDE) models. These two models,
combination of $\dot{H}$ and $H^2$, gave the late time accelerating
universe.

In \cite{48}, similar type of investigation has been done in FRW
universe model. In a recent paper \cite{49}, we have checked the
validity of the first and GSLT for Bianchi I universe model. We have
also explored the statefinder, deceleration and Hubble parameters
for the same line element \cite{50}. Here we extend the work of
\cite{48} to KK universe model with the same scenario. In this
paper, we use KK universe in thermal equilibrium composed of DM and
DE with GHDE and GRDE models. The paper is designed as follows: In
section \textbf{2}, the density and pressure for GHDE/GRDE models
are calculated. Section \textbf{3} is devoted to check the validity
of the first and GSLT on the apparent horizon and also by taking
GHDE/GRDE as the MHDE/MRDE. In the last section, we summarize the
results.

\section{Density and Pressure for GHDE and GRDE models}

In this section, we evaluate energy density and pressure for GHDE as
well as GRDE models in KK universe. This universe model contains
4-dimensional Einstein field equations and the fifth dimension
satisfies the Maxwell field equations. This metric is the simple
generalization of the FRW metric to extend the range of observable
universe by increasing the dimensions of the universe. The line
element of KK model is given by
\begin{equation}\label{1}
ds^2=-dt^2+a^2(t)[\frac{dr^2}{1-kr^2}+r^2(d\theta^2+\sin^2\theta d
\phi^2)+(1-kr^2)d\psi^2],
\end{equation}
where $k$ denotes the curvature parameter having values $+1, 0$ and
$-1$ corresponding to open, flat and closed universe, respectively.
The energy-momentum tensor for perfect fluid is
\begin{equation}\label{2}
T_{\alpha \beta}=(P+\rho)V_{\alpha} V_{\beta}-g_{\alpha
\beta}P,\quad(\alpha, \beta=0, 1, 2, 3, 4),
\end{equation}
where $P,~\rho$ and $V_{\alpha}$ are the pressure of the fluid,
energy density and five velocity vector, respectively. We consider
that the fluid is a mixture of DM and DE, thus $P$ and $\rho$ can be
written as $P=P_m+P_{E}$ and $\rho=\rho_m+\rho_{E}$ with $P_m=0$.
The field equations for KK universe become
\begin{eqnarray}\label{3}
8\pi\rho&=&6\frac{\dot{a}^2}{a^2}+6\frac{k}{a^2},\\\label{4}
8\pi
P&=&-3\frac{\ddot{a}}{a}-3\frac{\dot{a}^2}{a^2}-3\frac{k}{a^2}.
\end{eqnarray}

We are interested in flat KK universe so that $k=0$ yields the field
equations as
\begin{eqnarray}\label{5}
8\pi \rho=6\frac{\dot{a}^2}{a^2}=6H^2,\\\label{6}
8\pi
P=-3\frac{\ddot{a}}{a}-3\frac{\dot{a}^2}{a^2},
\end{eqnarray}
where Hubble parameter is defined as $H=\frac{\dot{a}}{a}$. The
conservation equation can be written as
\begin{equation}
\dot{\rho}+4H (\rho+P)=0.\label{7}
\end{equation}
Differentiating Eq.(\ref{5}) and using (\ref{7}), it follows that
\begin{equation}\label{8}
\dot{H}=-\frac{8\pi}{3}(\rho+P).
\end{equation}
Here we assume that there does not exist any sort of interaction
between DE and DM, therefore these are separately conserved. Thus
the conservation equation (\ref{8}) yields
\begin{eqnarray}\label{9}
\dot{\rho}_m+4H \rho_m=0,\\\label{10}
\dot{\rho}_{E}+4H(\rho_{E}+P_{E})=0.
\end{eqnarray}
Solving Eq.(\ref{9}), the matter energy density is obtained as
\begin{equation}\label{10*}
\rho_m=\rho_{m_0}(1+z)^4,
\end{equation}
where $\rho_{m_0}$ is the constant of integration, known as the
present value of DE density and cosmological red shift is
$z=\frac{1}{a}-1$. The matter density in KK universe decreases more
rapidly as compared to FRW universe with the evolution of the
universe which is consistent with the current observations.

Now, we evaluate energy density and pressure for GHDE and GRDE
models as follows.

\subsection{Generalized Holographic Dark Energy Model}

The energy density of this model is given as \cite{47}
\begin{equation}\label{11}
\rho_h=\rho_{E}=\frac{3c^2H^2}{8\pi}g\left(\frac{R}{H^2}\right),
\end{equation}
where $c^2$ is a non-zero arbitrary constant and $f(x)>0$ such that
$g(x)=\gamma x+(1-\gamma),~0\leq\gamma\leq1$. The Ricci scalar is
\begin{equation}\label{12}
R=-4(2\dot{H}+5H^2).
\end{equation}
For $\gamma=0$, the energy density of the original HDE is recovered
while $\gamma=1$ leads to the original RDE. Using Eq.(\ref{12}) in
(\ref{11}), it follows that
\begin{equation}\label{13}
\rho_h=\frac{3c^2}{8\pi}\left[(1-21\gamma)H^2-8\gamma
\dot{H}\right].
\end{equation}
Inserting Eqs.(\ref{10*}) and (\ref{13}) in (\ref{5}), we obtain a
first order linear differential equation whose solution is
\begin{equation}\label{14}
H^{2}=-\frac{8\pi\rho_{m_0}}{3}\frac{(1+z)^4}{((1+11\gamma)c^2-2)}+
H_{0}^{2}(1+z)^{\frac{2+c^2(21\gamma-1)}{8\gamma c^2}},
\end{equation}
with $H_0$ an integration constant. Differentiating Eq.(\ref{14})
with respect to $t$, we get
\begin{equation}\label{15}
\dot{H}=-\frac{16\pi\rho_{m_0}}{3(2-(1+11\gamma)c^2)}(1+z)^4-H_{0}^{2}
\frac{(2-(1-21\gamma)c^2)}{16\gamma
c^2}(1+z)^{\frac{2+c^2(21\gamma-1)}{8\gamma c^2}}.
\end{equation}
Substituting $H^2$ and $\dot{H}$ in Eqs.(\ref{10}), (\ref{12}) and
(\ref{13}), it follows that
\begin{eqnarray}\label{16}
P_h&=&-\frac{3H_{0}^{2}}{16\pi}\frac{((21\alpha-1)c^2+2)
((1-13\gamma)c^2-2)}{8\gamma
c^2}(1+z)^{\frac{2+c^2(21\gamma-1)}{8\gamma c^2}},\\\label{18}
R&=&-\frac{32\pi\rho_{m_0}}{3(2-(1+11\gamma)c^2)}(1+z)^4
+H_{0}^{2}\frac{(2+(1+19\gamma)c^2)}{2\gamma
c^2}\\\nonumber&\times&(1+z)^{\frac{2+c^2(21\gamma-1)}{8\gamma
c^2}},\\\label{17}
\rho_h&=&\frac{(1-37\gamma)c^2\rho_{m_0}}{(-2+(1+11\gamma)c^2)}(1+z)^4
-H_{0}^{2}\frac{3(2+(1-21\gamma)c^2)}{16\pi}\\\nonumber
&\times&(1+z)^{\frac{2+c^2(21\gamma-1)}{8\gamma c^2}}.
\end{eqnarray}
Equations (\ref{16}) and (\ref{17}) represent pressure and energy
density in terms of red shift $z$.

\subsection{Generalized Ricci Dark Energy Model}

The energy density of GRDE model is \cite{47}
\begin{equation}\label{19}
\rho_r=\frac{3c^2R}{8\pi} h\left(\frac{H^{2}}{R}\right),
\end{equation}
where $h(y)=\delta y+(1-\delta)>0,~0\leq\delta\leq1$. For
$\delta=0$, the original energy density of the RDE is recovered
whereas $\delta=1$ leads to energy density of the original HDE.
Comparing Eqs.(\ref{11}) and (\ref{19}), we see that the GRDE
reduces to the GHDE and vice versa for $\delta=1-\gamma$. By
replacing $\gamma$ with $(1-\delta)$ in Eqs.(\ref{13})-(\ref{17}),
we obtain similar solutions for GRDE model. This implies that these
equations are also solutions of the GRDE model with
$\gamma=1-\delta$.

\section{First and Generalized Second Law of Thermodynamics}

Firstly, we discuss the validity of the first and GSLT on the
apparent horizon. For this purpose, we use the entropy given by
Gibb's law \cite{51,52}
\begin{equation}\label{20}
T_AdS_I=PdV+d(E_A),
\end{equation}
where $S_I,~V,~P,~E_A$ and $T_A$ are internal entropy, volume,
pressure, internal energy and temperature of the apparent horizon,
respectively. In FRW metric, the apparent horizon has the radius
\begin{equation}\label{21}
R_A=\frac{1}{\sqrt{H^2+\frac{\kappa}{a^2}}}.
\end{equation}
Here FRW metric contained in the KK universe is a subspace with
compact fifth dimension having similar properties of flat FRW
universe on the apparent horizon. The internal energy and volume in
extra dimensional system are
\begin{equation*}
E_A=\rho V, \quad V=\pi^2 L^4/2.
\end{equation*}
In flat geometry, the radius of the apparent horizon coincides with
Hubble horizon given as
\begin{equation}\label{21+}
R_A=L=R_H=\frac{1}{H}.
\end{equation}
The entropy and temperature of the apparent horizon are \cite{53}
\begin{eqnarray}\label{22}
S_A=S_h=\frac{A}{4G},\quad (G=1),\quad T_A=\frac{1}{2\pi
R_A}=\frac{1}{2\pi L},
\end{eqnarray}
and entropy in four dimensions takes the form
\begin{eqnarray}\label{22+}
A=2\pi^2 L^3, \quad S_A=\frac{2\pi^2L^3}{4}=\frac{\pi^3 L^3}{2}.
\end{eqnarray}

The first law of thermodynamics on the apparent horizon is defined
as
\begin{equation}\label{23}
-dE_A=T_AdS_A.
\end{equation}
The energy crossing formula on the apparent horizon for KK universe
can be found as follows \cite{54}
\begin{eqnarray}\label{24}
-dE_A=2\pi^2R_A^4HT_{\alpha \beta}K^{\alpha}K^{\beta}dt
=2\pi^2R_A^4H(\rho+P)dt=-\frac{3\pi}{4}H\dot{H}L^4dt.
\end{eqnarray}
Inserting $L$ from Eq.(\ref{21+}) in the above equation, we get
\begin{equation}\label{25}
-dE_A=-\frac{3\pi}{4}\left(\frac{\dot{H}}{H^{3}}\right)dt.
\end{equation}
Using Eqs.(\ref{22}) and (\ref{22+}), it follows
\begin{equation}\label{26}
T_AdS_A=\frac{3\pi^2}{4}L\dot{L}dt=
-\frac{3\pi^2}{4}\left(\frac{\dot{H}}{H^{3}}\right)dt.
\end{equation}
These two equations lead to the following form of the first law of
thermodynamics
\begin{equation}\label{27}
-dE_A=\frac{1}{\pi}T_AdS_A,
\end{equation}
which gives its validity on the apparent horizon for all kinds of
energies as it is independent of DE.

Now for the GSLT to be satisfied for the apparent horizon, we
evaluate the derivative of internal entropy through Eq.(\ref{20}) as
\begin{equation}\label{27*}
\dot{S}_I=\frac{(\rho+P)\dot{V}+V\dot{\rho}}{T_A}.
\end{equation}
Substituting the values of $\dot{V},~T_A,~\dot{\rho}$ and using
conservation equation, we get
\begin{equation}\label{27+}
\dot{S}_I=\frac{3\pi \dot{H}R_A ^3(\dot{R_A}-R_A H)}{4T_A}.
\end{equation}
According to SLT, entropy of the thermodynamical system can never be
decreased. This is generalized in such a way that the derivative of
any entropy is always increasing, i.e., $\dot{S}_I+\dot{S}_A\geq 0$.
Thus we have
\begin{eqnarray}\label{28}
\dot{S}_I+\dot{S}_A=
\frac{3\pi^2}{8}\left[4\frac{\dot{H}^2}{H^6}-3\frac{\dot{H}}{H^4}\right]dt
\geq 0.
\end{eqnarray}
We conclude that GSLT always holds on the apparent horizon. Notice
that these laws always hold on the apparent horizon as it is
independent of choice of DE .

Further, we take GHDE or GRDE models as the density of MHDE or MRDE
to check the validity of the first and GSLT on the horizon having
radius $L$. The MHDE density can be calculated by taking the mass of
$(N+1)$-dimensional BH \cite{41}
\begin{equation*}
M=\frac{(N-1)A_{N-1}{R_H}^{N-2}}{16\pi G},
\end{equation*}
where $A_{N-1}$ is the unit $N$-sphere area, $R_H$ is the scale of
the BH horizon and $G$ is the gravitational constant in
$(N+1)$-dimensions related to Planck mass $M_{N+1}$. As $ 8\pi
G=M_{N+1}^{-(N-1)}=\frac{V_{N-3}}{M^2_p},~V_{N-3}$ is the volume
of this space, so $M$ can be written as
\begin{equation*}
M=\frac{(N-1)A_{N-1}{R_H}^{N-2}{M_p}^2}{2V_{N-3}}.
\end{equation*}
We can write
\begin{equation*}
L^3 \rho_\Lambda\sim\frac{(N-1)A_{N-1}{L}^{N-2}{M_p}^2}{2V_{N-3}},
\end{equation*}
which implies that
\begin{equation*}
\rho_\Lambda=\frac{c^2(N-1)A_{N-1}{L}^{N-5}{M_p}^2}{2V_{N-3}}.
\end{equation*}

For $N=4$, it gives
\begin{equation*}
\rho_\Lambda=\frac{3c^2A_3{L}^{-1}}{2}.
\end{equation*}
Inserting the value of 4-sphere area, we have
\begin{equation}\label{29}
\rho_\Lambda=\frac{3c^2\pi^2}{8}L^{2}.
\end{equation}
Comparing this value with the energy density of GHDE, it follows
that
\begin{equation}\label{30}
L^2=\gamma R+(1-\gamma)H^{2}.
\end{equation}
Substituting the values of $H^2$ and $R$ from Eqs.(\ref{14}) and
(\ref{18}) in (\ref{30}), the expression for $L^2$ in the form of
red shift is
\begin{eqnarray}\label{31}
L^2&=&\frac{1}{6c^2(-2+(1+11\gamma)c^2)}\times\left[-16\pi c^2
\rho_{m_0}(1-5\gamma)(1+z)^4\right.\nonumber\\&+&\left.3H_{0}^{2}
(-2+(1+11\gamma)c^2)(2+(1-21\gamma)c^2)(1+z)^{\frac{2+c^2
(21\gamma-1)}{8\gamma c^2}}\right].
\end{eqnarray}
Here the temperature, entropy and the total energy crossing on this
horizon with radius $L$ is similar to Eqs.(\ref{22}) and (\ref{24}),
respectively, with the difference that $dS_L,~T_L$ and $dE_L$ are
used instead of $dS_A,~T_A$ and $dE_A$. We can write
\begin{equation}\label{34}
T_L dS_L=\frac{3\pi^2}{4}L\dot{L}dt.
\end{equation}

For the first law, we must have $-dE_L=T_LdS_L$. Equations
(\ref{24}) and (\ref{34}) imply
\begin{equation}\label{35}
-dE_L=T_LdS_L-\frac{3\pi}{4}L\left[H\dot{H}L^3+\pi\dot{L}\right]dt.
\end{equation}
Since the second term on the right hand side in the above equation
is time dependent, so it can never be zero in the evolving universe.
Thus
\begin{equation}\nonumber
-dE_L \neq T_L dS_L.
\end{equation}
This indicates that the first law of thermodynamics does not hold
for the horizon of radius $L$. For the validity of GSLT on the
horizon of radius $L$, the derivative of total entropy is as follows
\begin{equation}\label{37}
\dot{S}_I+\dot{S}_L=\frac{3\pi^2}{8}{L}^2[4\dot{H}{L}^2(HL-\dot{L})+\pi
\dot{L}]dt.
\end{equation}
\begin{figure}
\centering\epsfig{file=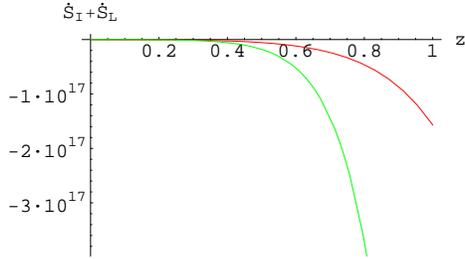, width=0.45\linewidth}\caption{The
graph shows the change of $(\dot{S}_I+\dot{S}_L)$ versus red shift
$z$ for $c=0.5,~\rho_{m_0}=1,~\gamma=0.7,~\delta=0.7,~H_0=70$. The
red colour represents GHDE and green represents GRDE.}
\end{figure}
According to the GSLT, the total entropy of the thermodynamical
system always increases, i.e., $4\dot{H}{L}^2(HL-\dot{L})+\pi
\dot{L}\geq 0$ indicating its dependence only on $L$ in the DE
model. In GHDE model, the variation of total entropy on the horizon
is
\begin{eqnarray}\label{38}
\dot{S}_I+\dot{S}_L=\frac{3}{8}\pi\left[4\dot{H}{L}^2(2L^2+(1+z)\frac{dL^2}{dz})
-\pi(1+z)\frac{dL^2}{dz}\right],
\end{eqnarray}
where $L^2$ is given in Eq.(\ref{31}). This expression does not
provide any indication, whether it increases or decreases. To get
insights, we plot a graph of total entropy $(\dot{S}_I+\dot{S}_L)$
versus red shift $z$ as shown in Figure \textbf{1}. This indicates
that $(\dot{S}_I+\dot{S}_L)<0$ and hence the GSLT does not hold on
this horizon with radius $L$ for the specific values of the
parameters.

\section{Concluding Remarks}

We have considered KK universe in compact form in the state of
thermal equilibrium, similar to FRW universe by assuming that our
universe is filled with DM and DE. Two types of DE models, GHDE and
GRDE have been used. It is worth noticing that the GRDE model can be
converted to GHDE model by interchanging $\delta$ with $1-\gamma$.
Also, the original density of HDE and RDE models is obtained for
$\gamma=0,~\delta=1$ and $\gamma=1,~\delta=0$, respectively. The
density and pressure for GHDE and GRDE models in terms of red shift
$z$ are evaluated.

We have investigated the validity of the first and GSLT on the
apparent horizon in this scenario. These laws turn out to be
independent of the choice of DE models, geometry of BH, and also the
fifth dimension. Hence these laws are always satisfied on the
apparent horizon for all kinds of DE models. We have also checked
that the first law remains invalid on particle as well as on the
event horizon while GSLT holds on the particle horizon only. It is
worth mentioning here that KK universe in non-compact form also
gives the same results as the variation along the fifth dimension is
negligible. Further, we have considered the GHDE and GRDE as the
MHDE and MRDE and found $L$ in terms of $z$, to check the validity
of these laws on the horizon whose radius is denoted by $L$. It is
concluded that the first law of thermodynamics does not hold on the
horizon of radius $L$ for both DE models. The GSLT always holds on
this horizon in the range $\gamma,~\delta\in(0,0.1)$ for GHDE and
GRDE, respectively, but it remains invalid for
$\gamma,~\delta\in(0.1,1)$. It is worth mentioning here that our
results on the apparent as well as on the horizon of radius $L$ are
consistent with FRW universe \cite{48}.

\end{document}